\documentclass{article}
\usepackage{spconf,amsmath, amssymb, graphicx}
\usepackage{algorithm}
\usepackage{subcaption}
\usepackage{multirow}
\usepackage[noend]{algpseudocode}

\title{Exploring Federated Self-Supervised Learning for General Purpose
Audio Understanding}
\name{Yasar Abbas Ur Rehman, Kin Wai Lau, Yuyang Xie, Lan Ma, Jiajun Shen }
\address{TCL AI Lab, Hong Kong}
\begin{document}
\topmargin=0mm 
%
\maketitle
\begin{abstract}
The integration of Federated Learning (FL) and Self-supervised Learning (SSL) offers a unique and synergetic combination to exploit the audio data for general-purpose audio understanding, without compromising user data privacy. However, rare efforts have been made to investigate the SSL models in the FL regime for general-purpose audio understanding, especially when the training data is generated by large-scale heterogeneous audio sources. In this paper, we evaluate the performance of feature-matching and predictive audio-SSL techniques when integrated into large-scale FL settings simulated with non-independently identically distributed (non-iid) data. We propose a novel Federated SSL (F-SSL) framework, dubbed FASSL, that enables learning \textit{intermediate} feature representations from large-scale decentralized heterogeneous clients, holding unlabelled audio data. Our study has found that audio F-SSL approaches perform on par with the centralized audio-SSL approaches on the audio-retrieval task. Extensive experiments demonstrate the effectiveness and significance of FASSL as it assists in obtaining the optimal global model for state-of-the-art FL aggregation methods. 



\end{abstract}

\begin{keywords}
Audio understanding, Self-supervised learning, Federated Learning
\end{keywords}
%
\section{Introduction}
Self-supervised learning (SSL) enables learning of \textit{intermediate} audio representations from unlabeled audio contents, which are used to solve various downstream tasks e.g., automatic speech recognition, speaker recognition, audio event localization \cite{wang2022towards, wei2021inferring, yang2020soundr, ahuja2020direction}. However, the utility of audio-SSL in its naïve form requires the sensitive audio data to be sent to a centralized server for processing, posing serious concerns around privacy, increasing communication and storage costs, and ultimately limiting the scope of the technology to small-scale audio datasets \cite{villalobos2022will, zhuang2023foundation}.

To overcome these limitations of centralized audio-SSL,  the combination of Federated Learning (FL) techniques and audio-SSL proposes an alternative solution. FL is a decentralized distributed learning framework in which a distributed population of edge devices collaboratively trains a machine learning model while keeping the data secure and private. FL combined with SSL (F-SSL) offers a unique decentralized learning framework that overcomes the limitations of both SSL and FL. For example, F-SSL enables large-scale decentralized feature learning without requiring laborious data annotations and has been shown to provide competitive performance against centralized SSL, sometimes even surpassing it on certain downstream tasks.  \cite{gao2022end,rehman2022federated, rehman2023ldawa}. 

Numerous studies have explored the F-SSL framework to learn feature representations from audio data \cite{feng2022federated, gao2022federated, saeed2020federated, zhang2023fedaudio}. However, they overlook the following prevalent issues in audio F-SSL. 
(1) The existing approaches for \textit{general-purpose audio F-SSL} lack a side-by-side comparison of feature-matching and predictive SSL approaches in the \textit{cross-device} FL setting.   
(2) Current methods for general-purpose audio F-SSL only adopt the most basic federated learning algorithm, i.e., FedAvg \cite{mcmahan2017communication}, leaving a large space on the table for performance improvement and further exploration. 
To resolve these gaps for \textit{general-purpose} audio understanding, 
we conduct the first comprehensive study on audio F-SSL in the \textit{cross-device} FL settings with non-iid audio data.  We note that in the \textit{cross-device} FL setup, it is quite perplexing to reach any conclusive evidence about the superiority of one audio-SSL approach over another audio-SSL approach across the heterogeneous downstream tasks (e.g., semantic audio understanding vs non-semantic audio understanding). Subsequently, the same is true for the F-SSL aggregation methods. 

 Based on the findings of our study of audio-SSL with FL, we propose a novel FL framework, FASSL, designed specifically for audio F-SSL. This framework allows the server to dynamically allocate and update the global model for optimal performance on the range of heterogeneous audio downstream tasks. Using performance metrics as the radar, FASSL can identify when the global model is optimal and suboptimal for a certain audio downstream task while executing F-SSL training. Our framework is agnostic to the type of SSL and FL aggregation methods.  The main contribution of this work is as follows: (1) We provide a first systematic study of audio F-SSL (both contrastive and non-contrastive) pretext-tasks for learning \textit{general-purpose} audio representations from decentralized non-iid data in \textit{cross-device }settings.
(2) We perform in-depth analyses of a variety of image-based/video-based F-SSL aggregation methods for audio-representation learning, which remained unexplored previously.  
(3)We propose a framework, dubbed FASSL, which enables the server to identify the optimal global model for the audio downstream tasks during FL pretraining of audio-SSL approaches. 

\section{Methodology}

\begin{figure}[t]
    \centering
    \includegraphics[width=1\linewidth, height=0.4\linewidth]{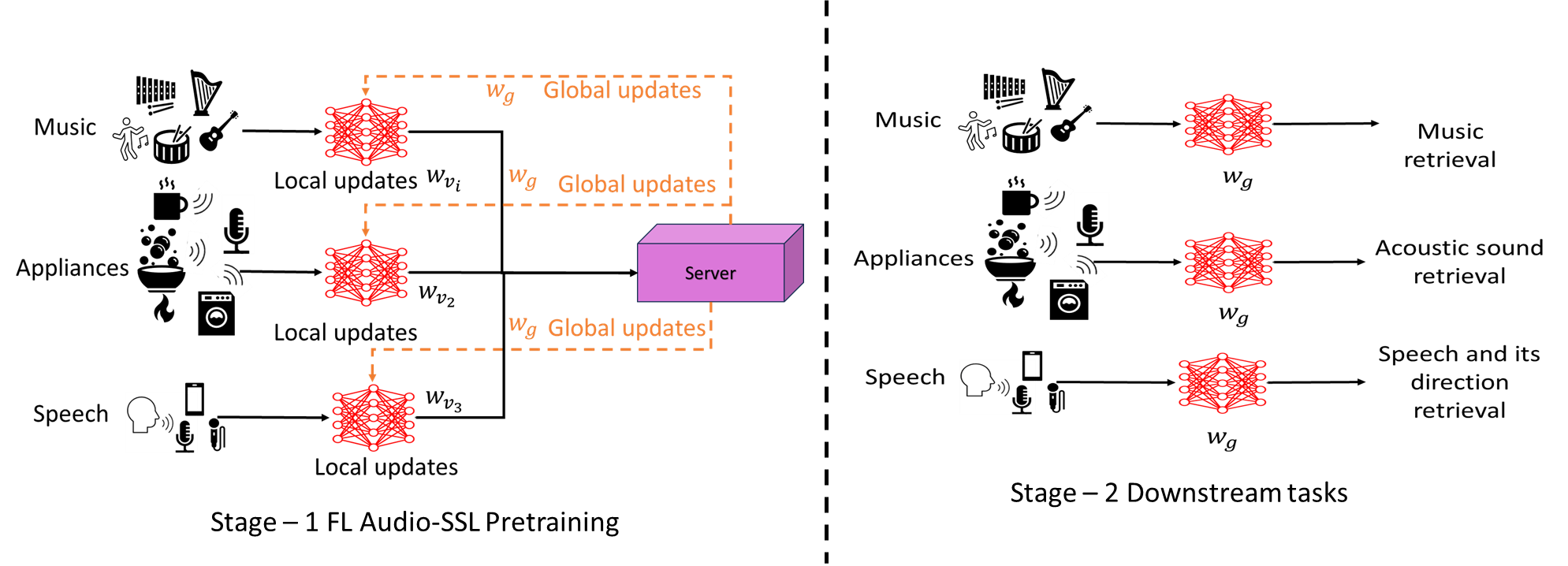}
    \vspace{-5mm}
    \caption{System overview: Stage-1 (left) represents the FL audio-SSL pertaining. The downstream task; audio retrieval is depicted in Stage 2 (right). Note that the data of stage-1 and stage-2 are from different sources}
    \label{fig:f-ssl-setup}
\end{figure}

We adopted the procedure of \cite{rehman2022federated} to train the audio-SSL method in FL. Specifically, during each federated round $r$  ($1 \leq r \leq R$), the server sends the global model $w_g$ to a set of randomly selected clients $v_{i}$. Each of the selected clients then updates its local model by running the SSL pretext task (contrastive or non-contrastive) on their local data for $E$ number of epochs with $w_g$ as an initialization.
At the end of the $E$ number of local epochs, the client sends the updated model to the server for aggregation. The server aggregates all the incoming models using the aggregation strategy to generate a new global model $w_g$ for the next round. This has been shown in Fig. \ref{fig:f-ssl-setup}. The aggregation strategy in general can be defined as follows: $w_{g}^{r+1} =  \sum_{i=1}^{s} \beta_{i} * w^{r}_{v_{i}}$, where, the formulation for $\beta$ depends on the aggregation strategy. 



\noindent\textbf{Optimal Global Model Selection for Downstream Task:}
After completing local pretraining at round $r$, the participating clients' locally updated weights $w_{v_{i}}$ are collected by the server. The server then aggregates the local models according to a predefined strategy and generates a new global model $w^{r+1}_{g}$ for the next round of FL followed by evaluating the new global model on the set of heterogeneous downstream tasks. If the performance, top-1\% accuracy here,  of the updated global model $w^{r+1}_{g}$ on the $ith$ downstream task $dt_{i}$ gets higher than the previous optimal global model $w^{dt_{i}}_{g_{opt}}$ for the $dt_{i}$, the server simply replaces $w^{dt_{i}}_{g_{opt}}$ with $w^{r+1}_{g}$, otherwise, the $w^{dt_{i}}_{g_{opt}}$ remains unchanged. 

\begin{equation}
  w^{dt_{i}}_{g_{opt}} = 
  \begin{cases} 
       w_{g}^{r+1} & if ~~acc(w_{g}^{r+1}) > acc( w^{dt_{i}}_{g_{opt}}) \\
       w^{dt_{i}}_{g_{opt}} & Otherwise \\
       
   \end{cases}
\end{equation}

One can use other performance metrics instead of accuracy. However, we limit ourselves to the top-1\% retrieval accuracy. In this work, we considered several FL aggregation methods, i.e.,  FedAvg \cite{mcmahan2017communication}, FairAvg\cite{michieli2021all}, Loss\cite{gao2022end}, FedU \cite{zhuang2021collaborative}, and L-DAWA \cite{rehman2022federated}, although theoretically, any number of aggregation methods are possible.

\subsection{Implementation details}
\textbf{Pretraining:}
We consider three SSL techniques in this work, i.e., ACOP \cite{xu2019self}, SimCLR\cite{chen2020simple}, and Barlow Twins \cite{zbontar2021barlow}. We use ResNet-18 \cite{he2016deep} as the backbone network for all the SSL tasks. 
For the FL pretraining of audio-SSL methods, unless otherwise specified, the local pretraining on each client lasts for $E=1$ epoch per FL round $R$. We set the total number of rounds $R$ to $100$ to ensure that each client acquires sufficient participation during FL pretraining. For each round, we randomly select 10 clients from the pool of 100 clients to participate in training. Each selected client then trains its local model using an SGD optimizer with similar parameters as centralized SSL except that we maintain a constant learning rate throughout the training. We set the batch size to 64. All the experiments were performed using the Flower \cite{beutel2020flower} framework with PyTorch Lightning \cite{falcon2019pytorch}.

\noindent\textbf{Downstream Task Evaluation:}
We follow the protocol of \cite{xu2019self} to report the Top-1 \% audio retrieval performance. The features of the samples in the test audio datasets are used as a query to find the corresponding features of the samples in the training audio datasets. When the class of the test samples appears in the class of $k$ nearest training samples, it is said to be correctly predicted.     

\noindent\textbf{Pretext Task Dataset:}
We pretrain the SSL methods considered in this work on the VGG Sound dataset \cite{chen2020vggsound}. It is a large-scale audio dataset extracted from YouTube videos with more than 200K audio clips, each is 10 seconds long sampled at 16KHz. There are 309 classes in total. 

\noindent\textbf{VGG Sound Non-iid:} To simulate a realistic FL environment, we generate non-iid versions of VGG Sound based on actual class labels using a Dirichlet coefficient $\alpha$, where a lower value indicates greater heterogeneity \cite{lubana2022orchestra, zhang2023fedaudio}. In this paper, we set the value of $\alpha$ to 0.1. As a result, the datasets are randomly partitioned into 100 shards for \textit{cross-device} setting to mimic the setup of having 100 disjoint clients participating in FL \cite{rehman2022federated}.  



\noindent\textbf{Downstream Task Dataset:}
For the downstream tasks we use Speech Commands V2 (KS2) \cite{speechcommandsv2}, Direction of Voice  (DOV) \cite{ahuja2020direction}, EPIC-Sound \cite{huh2023epic}, and NSYNTH datasets \cite{engel2017neural}. 
\section{Experiments}


\begin{table*}[h!]
\centering
\resizebox{2\columnwidth}{!}{
    \centering
    \begin{tabular}{c|llll|cc|cc|cc|cc}
    \hline
          & \multicolumn{4}{c|}{Predictive} & \multicolumn{8}{c}{Feature Matching}\\ 
          \cline{2-13}
         Task   & KS2 & DOV & Epic-Sound & NSYNTH &    \multicolumn{2}{|c}{KS2} & \multicolumn{2}{|c}{DOV} & \multicolumn{2}{|c}{EPIC-Sound} &  \multicolumn{2}{|c}{NSYNTH}   \\
         \hline
        Method  & \multicolumn{4}{c|}{ACOP} &   BT & SimCLR  &  BT & SimCLR  & BT & SimCLR  & BT & SimCLR  \\
         \hline
         Centralized  & 12.39 &66.12&  18.02&  60.96& 10.06 & 10.26 & 66.77 & 66.30 & 25.37 & 26.10 & 69.58 & 68.55  \\
         Federated (vanilla) & 12.61 &65.00& 18.46 & 60.66 &  12.70  & 15.82 & 69.06 & 69.40 &
          21.51 & 19.98 & 61.52 & 60.32 \\
          \hline
          FASSL & 12.92 (1) & 66.33 (1) & 19.55 (90) & 60.83 (90) &  18.50 (1) & 16.28 (40) & 69.35 (60) & 69.49 (80) & 21.72(90) & 19.98 (100) & 61.52 (100) & 61.23 (1) \\
         \hline
    \end{tabular}}
    \vspace{-3mm}
    \caption{Audio retrieval performance (Top-1\%) for three audio-SSL methods in centralized, federated, and FASSL settings. The number in the parenthesis "()" indicates the FL rounds where the global model performance is optimal for the downstream task.}
    \label{tab:Fl_ssl}
\end{table*}

\subsection{Centralized vs. Federated Audio-SSL}
Table \ref{tab:Fl_ssl} shows the comparison between centralized audio-SSL and its FL counterpart. First, the FL version of ACOP has obtained comparable or better performance than its centralized counterpart across all the downstream tasks. This observation is in line with findings of \cite{rehman2022federated}, which shows that the FL pretraining of predictive SSL methods results in a wider-loss landscape than their centralized counterpart and hence better performance. Second, we found that the FL versions of SimCLR and Barlow Twins have obtained better performance on KS2 and DOV datasets while lower performance on the EPIC-Sound and NSYNTH datasets. Third, the SimCLR and Barlow Twins provide better transfer learning performance than ACOP both in the centralized and FL settings. One can also see from  Table \ref{tab:Fl_ssl} that the average performance of vanilla F-SSL methods is on par with centralized SSL methods.

Interestingly, FASSL identifies that ACOP is unable to learn a better global model for KS2 and DOV after round 1 while achieving a better model for EPIC-Sound and NSYNTH at round 90. FASSL also found that the BT does not perform better on KS2 after round 1 while SimCLR is unable to learn a better model on NSYNTH after round 1. These results also show how intermediate model evaluation can identify whether the F-SSL training is correlated with the downstream task.       

\subsection{Effects of FL Aggregation Methods}
In this experiment, we investigate the impact of FL aggregation strategies on the pretraining of the audio-SSL approaches in Non-iid \textit{cross-device} FL settings. We evaluated several aggregation schemes that are proposed for F-SSL. These include Loss \cite{gao2022end}, FairAvg\cite{michieli2021all}, FedU \cite{zhuang2021collaborative}, and L-DAWA\cite{rehman2023ldawa}. One can see from Table \ref{tab:FM-Aggreg} that the performance of all aggregation methods is similar across the downstream tasks. More importantly, FASSL identifies that for certain downstream tasks, the FL aggregation methods are unable to learn a better model throughout the training. This motivates us to investigate whether excluding certain parts of the model from being transceived by the server and clients can provide any better performance across the downstream tasks, which we explored in the following section.

\begin{table*}[h!]
\centering
\resizebox{2\columnwidth}{!}{
    \centering
    \begin{tabular}{l|llll|ll|ll|ll|ll}
    \hline
         Task   & KS2 & DOV & Epic-Sound & NSYNTH &  \multicolumn{2}{c}{KS2} & \multicolumn{2}{|c}{DOV} & \multicolumn{2}{|c}{EPIC-Sound} &  \multicolumn{2}{|c}{NSYNTH}   \\
         \hline
        Method  & \multicolumn{4}{c}{ACOP} &  BT & SimCLR  &  BT & SimCLR  & BT & SimCLR  & BT & SimCLR  \\
         \hline
          FedAvg & \textbf{12.61} & 65.00 & 18.46 & 60.66 &  12.70  & \textbf{15.82} &\textbf{ 69.06} & 69.40 & 21.51 & 19.98 & 61.52 & \textbf{60.32} \\
          FairAvg & 12.18 & 64.70& 18.70& \textbf{60.76}&   13.26 & 15.57 & 67.56  & 69.77 &  22.06 & 19.91 & 61.84 & 59.44  \\
          Loss & 12.15 & 65.23 & 18.50 & 60.47 & 13.61  & 15.61  &  68.51 & 69.61 & \textbf{22.22} & \textbf{20.20} & 60.93 &  59.79 \\
          FedU & 12.52& \textbf{65.71}& 18.30 & 60.44&  13.12 & 15.67  &  68.73 & 69.62& 21.57 & 19.91 & 60.74 &  60.03 \\
          L-DAWA &  12.22 &  65.21 & \textbf{18.75}  &  59.93 &  \textbf{13.73}  &  15.68 &  67.15 & \textbf{69.82}& 21.85 & 20.01 & \textbf{61.86 }&  59.79 \\
          \hline
        FASSL-FedAvg &  12.92 (1) & 66.33 (1) & \textbf{19.55} (90) & 60.83 (90) & 18.50 (1) & 16.28 (40) & 69.35 (60) & 69.49 (80) & 21.72(90) & 19.98 (100) & 61.52 (100) & \textbf{61.23} (1) \\
        FASSL-FairAvg & 12.59 (1)& 65.25 (1)& 19.01 (60)&61.10 (50) & 17.21 (1)  & \textbf{16.40} (50) & 69.03 (70) & 69.77 (100) & 22.06 (100) & 19.91 (100) & 61.84 (100) &  60.88 (1)\\
        FASSL-Loss & 12.45 (1) & 65.71 (1) & 18.96 (1) & 60.86 (50) & \textbf{18.59 }(1) & 16.36 (50)& 69.32 (1) & 69.66 (90) & \textbf{22.22} (100) & \textbf{20.20} (100) & 61.32 (80)&  60.98 (1) \\
        FASSL-FedU & \textbf{12.96} (10) & \textbf{65.71} (100)& 19.48 (1)& \textbf{60.98 }(20) & 18.13 (1) & 16.30 (50)& \textbf{69.54} (40) & \textbf{69.83} (70) & 21.95 (90) &  20.01 (90) & 60.74 (100)&  61.27 (1) \\
        FASSL-L-DAWA & 12.92 (10) & 65.48 (60) & 19.51 (1)  & 60.54 (1)& 17.68 (1) &  16.11 (50) & 69.17 (1) & \textbf{69.82} (100) & 21.95(90) &  20.01(100) & \textbf{61.86 }(100) & 60.76 (1)   \\

         \hline
    \end{tabular}}
    \vspace{-2mm}
    \caption{Audio retrieval performance (Top-1\%) of feature-matching audio-SSL methods in cross-device FL settings. The number in the parenthesis "()" indicates the FL rounds where the global model performance is optimal for the downstream task.}
    \label{tab:FM-Aggreg}
\end{table*}





\subsection{Transceiving only the backbone weights}
In \cite{rehman2022federated}, the authors suggested that only transceiving (transmit and receive) the backbone weights in FL can improve performance. They argue that the classification head is more representative of the local client data distribution, which can lead to higher divergence among the weights during aggregation. In contrast to \cite{rehman2022federated}, we investigate this phenomenon for FedAvg, Loss, FairAvg, and L-DAWA in the context of audio F-SSL. 

One can see from Table \ref{tab:FM-Aggreg-bb} that only transceiving the backbone (aggregation method with the postscript "-bb"), in general, can improve the audio retrieval performance of the F-SSL-based model across all the downstream audio tasks except for NSYNTH. With FASSL, we further found that the ACOP, with only transceiving the backbone weights, learns better feature representations as the training progresses. 
However, the optimal performance with FASSL for BT and SimCLR has slightly degraded compared to the full model results. 

\begin{table*}[h!]
\centering
\resizebox{2\columnwidth}{!}{
    \centering
    \begin{tabular}{l|llll|ll|ll|ll|ll}
    \hline
         Task   & KS2 &DOV&  EPIC-Sound&  NSYNTH & \multicolumn{2}{c}{KS2} & \multicolumn{2}{|c}{DOV} & \multicolumn{2}{|c}{EPIC-Sound} &  \multicolumn{2}{|c}{NSYNTH}   \\
         \hline
        Method  & \multicolumn{4}{c}{ACOP} &  BT & SimCLR  &  BT & SimCLR  & BT & SimCLR  & BT & SimCLR  \\
          \hline 

        FedAvg-bb & 13.60 & 66.11 & 18.75 & 59.39 & 14.30 &  \textbf{15.89} & 68.42 & 69.77 &  21.68 & 19.52 & \textbf{61.32} & \textbf{60.25}\\
        FairAvg-bb &  13.04 & 65.96 & 19.17 & \textbf{59.91} &  13.84 & 15.83 & 68.98  & 69.84 & 21.77 & 19.91 & 60.79 &  59.59 \\
        Loss-bb &  \textbf{13.67} & \textbf{66.11} & \textbf{19.28} &  59.35 & 14.16 &   15.60 & 69.44  & 70.02 & \textbf{22.11} &\textbf{ 20.03} & 61.27 &  59.93 \\
        L-DAWA-bb & 13.43 &  65.76 & 18.81  & 59.59 & \textbf{14.57}  & 15.47 & \textbf{69.50}  &\textbf{ 70.17}&  21.45 & 20.00 & 60.71 & 60.10  \\
        \hline
        
        FASSL-FedAvg-bb &13.60 (100) & 66.25 (90) & 18.90 (10) &60.66 (10) & \textbf{18.18}(1)  &  \textbf{16.30} (50) &  69.47 (40) & 69.78 (90) & 21.68 (100) & 19.56 (70) & 61.32 (100) & \textbf{61.40} (1) \\
        FASSL-FairAvg-bb & 13.19 (90) & 65.96 (100) & 19.36 (40) & 60.66 (10)& 18.05 (1) & 15.97 (40) & 68.98 (100) & 69.84 (100) & 21.77 (100) & 19.91 (100) & 61.20 (90) & 61.18 (1)  \\
        FASSL-Loss-bb & \textbf{13.67} (100) &  \textbf{66.11} (100)& \textbf{19.56 }(90) & \textbf{60.91} (1)& 17.74 (1) & 16.02 (40) & \textbf{69.52} (80) & 70.02 (100) & \textbf{22.11} (100) &\textbf{ 20.03} (100) &\textbf{ 62.01} (80) &  60.91(1) \\ 
        FASSL-L-DAWA-bb & 13.62 (90) & 65.96 (1) & 19.21 (10)  &  60.49 (1) & 17.53 (1) &  15.93 (80) & 69.50 (100) &  \textbf{70.17} (100)& 21.45 (100) & 20.01 (60) & 60.71 (100)&  60.40 (90)  \\
          
         \hline
    \end{tabular}}
    \vspace{-2mm}
    \caption{Audio retrieval performance (Top-1\%) of feature-matching audio-SSL methods in cross-device FL settings. The number in the parenthesis "()" indicates the FL rounds where the global model performance is optimal for the downstream task.}
    \label{tab:FM-Aggreg-bb}
\end{table*}

\subsection{Effects of the Local Epochs}

As shown in Table \ref{tab:FM-Aggreg-bb-epochs}, the predictive approach, ACOP, with 10 local epochs and L-DAWA, provides an all-time best performance of 22.61\% after 100 rounds and 23.35\% as identified by FASSL at round 30. Except for NSYNTH at 10 local epochs, we found that increasing the local epochs to 10 helps ACOP to get better performance on DOV, Epic-Sound, and NSYNTH tasks.  We further note from the results in Table \ref{tab:FM-Aggreg-bb-epochs}, that with increasing local epochs, the performance of feature-matching approaches generally degrades on KS2 and improves on Epic-Sound and NSYNTH.  We conjecture that this is due to the large variations in the characteristics of semantic data (speech) compared to non-semantic data (music, event sounds) on the local clients, the ability of feature-matching approaches to learn more discriminative feature representations, and overfitting on the local data, which causes the clients' models to drift away from each other causing the global model to diverge from its optimization path by a large margin.

\begin{table*}[h!]
\centering
\resizebox{2\columnwidth}{!}{
    \centering
    \begin{tabular}{l|l|llll|ll|ll|ll|ll}
    \hline
          & Task   & KS2 & DOV & Epic-Sound & NSYNTH & \multicolumn{2}{c}{KS2} & \multicolumn{2}{|c}{DOV} & \multicolumn{2}{|c}{EPIC-Sound} &  \multicolumn{2}{|c}{NSYNTH}   \\
         \hline
        LE/R & Method  & \multicolumn{4}{c}{ACOP}  & BT & SimCLR  &  BT & SimCLR  & BT & SimCLR  & BT & SimCLR  \\
          \hline 
    \multirow{6}{*}{5} & FedAvg-bb & 15.53 & 67.45 & 18.56 & 60.44 & 14.36 & \textbf{12.07} & \textbf{67.93} & 68.61 & 22.58 & 23.11 & \textbf{64.03} & 62.40\\
    
    & Loss-bb & 14.76 & 68.27 & 18.12 & 60.59 & 15.56 & 11.82 & 67.06 & \textbf{69.24} &  \textbf{22.61} & 22.88 & 61.71 & 62.10 \\
    
    & FairAvg-bb & 14.35 & \textbf{68.37} & 17.91 & \textbf{61.84} & 14.55 & 11.80 & 65.80 & 68.77 & 21.80 & 23.08 & 61.47 & 61.93\\
    
    & L-DAWA-bb & \textbf{18.00} & 68.01 & \textbf{19.20} & 59.27 & \textbf{16.02} & 11.71 & 66.07 & 69.16 & 21.62 & \textbf{23.35} & 61.86 & \textbf{62.89} \\
    \cline{2-14}
    & FASSL-FedAvg-bb & \textbf{18.92 }(10) & 67.62(30) & 19.10 (70) & 60.71 (60) & \textbf{17.27} (70) & \textbf{14.13}(1) & \textbf{67.93} (100) & 68.79 (80) & 22.58 (100) & 23.11 (100) & \textbf{64.03} (100) & 62.71 (90) \\ 
    
    & FASSL-Loss-bb &14.77 (30) & 68.42 (10) & 18.76 (30) & 61.10 (60) & 15.56 (100)  & 13.61 (1) & 67.56 (80) & \textbf{69.24} (100)  & \textbf{22.61} (100) & 23.17 (90) &  61.71 (100)& 62.67 (90) \\   
    & FASSL-FairAvg-bb & 14.99 (80) & \textbf{68.94} (20) & 18.72 (40) & \textbf{62.03} (70) & 14.55 (100) & 14.09(1)  & 66.85 (40) &  69.01 (90) & 21.80 (100)  &\textbf{ 23.46} (80) & 61.47 (100) & 61.93 (100) \\
    & FASSL-L-DAWA-bb & 18.80 (10) & 68.86 (90) & \textbf{19.73} (10) & 60.05 (80) & 16.02 (100) & 13.98 (1) &  67.90 (10) & 69.16 (100) &  21.87 (90) & 23.35 (100) & 61.86 (100) & \textbf{62.89} (100) \\
    \hline 
\multirow{4}{*}{10}& FedAvg-bb &  16.09 & 66.84 & 19.16 & \textbf{60.96} & 13.75 & 10.67 & \textbf{69.32} & 69.18 & \textbf{25.03} & 23.17 & 62.37 & \textbf{64.28} \\
& Loss-bb & 13.66 & \textbf{68.27} & 18.31 & 59.86 & \textbf{15.29} & 10.80 & 68.40 &\textbf{69.42} & 24.76 & \textbf{23.85} &  64.06 & 63.28  \\
& FairAvg-bb & 17.02 & 67.70 & \textbf{19.96} & 60.13 &  15.23 & 10.70 & 67.88 & 69.09 & 24.06 & 23.81 & \textbf{65.33}  & 62.69  \\
& L-DAWA-bb & \textbf{22.61} & 65.32 & 19.58 & 57.17 & 11.95 & \textbf{10.91} & 67.75 & 69.10 &  24.78 & 22.97 & 61.62 & 63.59 \\
\cline{2-14}
& FASSL-FedAvg-bb & 18.80 (40) & 67.13 (60) & 19.25 (50) & \textbf{61.30} (1) & 14.54 (50) & 14.32 (1) &\textbf{69.32} (100) & 69.27 (90) & 25.11 (90) & 23.17 (100) & 64.57 (90) & \textbf{64.28} (100) \\
& FASSL-Los-bb & 13.95 (20) & \textbf{68.75 }(70) &  19.01 (20) & 61.20 (70) & \textbf{16.27} (90) & 13.87 (1) & 68.40 (100) & 69.42(100) & 24.76 (100) & \textbf{23.85} (100) & 64.08 (90) &  63.54 (90) \\
& FASSL-FairAvg-bb &  18.29 (20)& 68.14 (40)& \textbf{19.96} (100)& 60.18 (1) & 15.88 (90) & 13.89 (1)& 68.17 (80) & \textbf{69.45} (70) & 24.06 (100) & 23.81(100) & \textbf{65.33} (100) & 63.69 (90)\\
& FASSL-L-DAWA-bb & \textbf{23.35} (30) & 65.81 (80) & 19.87 (10) & 58.20 (1) & 15.59 (70) & 13.89 (1)  & 69.13 (90)& 69.31 (70) & \textbf{25.33} (90) & 23.52 (70)  & 64.52 (80)& 63.59 (100) \\
         \hline
    \end{tabular}}
    \vspace{-2mm}
    \caption{Audio retrieval performance (Top-1\%) of feature-matching audio-SSL methods in cross-device FL settings. The number in the parenthesis "()" indicates the FL rounds where the global model performance is optimal for the downstream task}
    \label{tab:FM-Aggreg-bb-epochs}
\end{table*}





\subsection{Discussion}

The observations made in the previous section suggest that predictive approaches learn better audio feature representations by transceiving only the backbone part of the model. Consequently, predictive approaches, such as ACOP, perform better on speech-related tasks. 
For a fixed number of communication rounds and with increasing local epochs, we found that ACOP provides better performance with divergence-based aggregation strategies such as L-DAWA. We further found that once the number of local epochs is increased, beyond 1 local epoch, ACOP performs much better than SimCLR and Barlow Twins on the KS2 dataset. 

For the feature-matching approaches (BT and SimCLR), we observe that they perform much better on the non-semantic audio understanding tasks. For example, in vanilla FL settings feature-matching approaches show improved performance with an increasing number of communication rounds. Although, in terms of optimal performance, FASSL identifies no benefit of transceiving the backbone layer with feature-matching approaches, its role became more prominent when the number of local epochs increased beyond 1. We further found that increasing the number of local epochs in feature-matching approaches showed improved performance across the aggregation methods on non-semantic audio understanding tasks as the F-SSL training progressed. This shows a promising direction for future exploration.      

\section{Conclusion}
 We proposed the first systematic study on pertaining audio-SSL methods in FL \textit{cross-device} settings with non-iid data. Subsequently, we proposed FASSL to find whether the FL pretraining of audio-SSL approaches is correlated with the downstream task and to identify the optimal global models for the heterogeneous downstream tasks. In our study, we found that the performance audio-SSL methods with vanilla FL, in \textit{cross-device} setup with non-iid data, surprisingly perform on par with their corresponding centralized counterparts. We further found that transceiving only the backbone part of the model during FL pretraining improves the audio retrieval performance. Interestingly, from the experimental results, we noted that the performance of contrastive SSL techniques improves on non-semantic audio understanding tasks while the predictive approaches show significant performance improvement on the semantic audio understanding tasks. We hope this work will enable future research toward the integration of FL with SSL for general-purpose audio understanding.         

\clearpage
\bibliographystyle{IEEEbib}
\bibliography{bibliography}

\end{document}